\documentclass[
    ,final            
  ]
  {aipproc}

\layoutstyle{8x11single}
\usepackage{axodraw}

\begin{document}

\title{Efficient Energy Transport in Photosynthesis:\\
       Roles of Coherence and Entanglement\footnote{
Invited talk presented at the Symposium ``75 Years of Quantum Entanglement:
Foundations and Information Theoretic Applications", January 2011, Kolkata,
India. AIP Conference Proceedings (2011).}}


\author{Apoorva D. Patel}{
  address={Centre for High Energy Physics and
           Supercomputer Education and Research Centre,\\
           Indian Institute of Science, Bangalore 560012, India}
}

\begin{abstract}
Recently it has been discovered---contrary to expectations of physicists
as well as biologists---that the energy transport during photosynthesis,
from the chlorophyll pigment that captures the photon to the reaction
centre where glucose is synthesised from carbon dioxide and water, is
highly coherent even at ambient temperature and in the cellular environment.
This process and the key molecular ingredients that it depends on are
described. By looking at the process from the computer science view-point,
we can study what has been optimised and how. A spatial search algorithmic
model based on robust features of wave dynamics is presented.
\end{abstract}

\maketitle


At the outset, I thank the organisers of this conference, for allowing me
to talk on a topic with little direct connection to the main theme of the
conference. Nevertheless, photosynthesis is a wonderful phenomenon of life
with crucial involvement of quantum effects, and its detailed understanding
is bound to teach us how to take advantage of quantum effects in practical
problems. In the following, I present a physicist's model of the problem,
sufficiently simplified for a precise mathematical treatment, and the key
features of its solution.

\section{The Problem} 

Life is a fundamentally non-equilibrium process. Living organisms need to
stay out of equilibrium in order to survive, prosper and reproduce, and
that requires a continuous supply of free energy. (When living organisms
come to equilibrium with their environment, we call them dead.) The primary
source of this free energy is the sun. Photosynthesis is the phenomenon
whereby the chlorophyll pigments of plants capture sunlight and convert
that energy in to a chemical form for later use. It is the basic step of
harnessing solar energy on which most living organisms depend, either
directly or indirectly through a food chain.

In the first step of photosynthesis, the energy of the captured photon
is used to dissociate water (in to $H^+$ and $OH^-$) and create charge
separation across a membrane. This is essentially an electrical step,
and the resultant ions subsequently drive chemical processes for the
synthesis of glucose. It is established that the efficiency of this
energy conversion exceeds 95\% (some quote even 99\%). One can easily
estimate the efficiency by counting the number of photons captured and
the number of glucose molecules synthesised, since energies of the
radiation and the chemical bonds are known. On the other hand, the
dynamics behind the high efficiency is still a mystery to be unraveled,
especially how the dissipative effects of the chaotic cellular environment
are kept out. See Ref.\cite{psReview} for a review.

Even though we do not know how nature arrived at the nearly perfect
optimisation of the energy conversion efficiency, we can argue that
biological evolution governed by Darwinian selection would have led to it.
In contrast, our best solar cells with the same aim (i.e. use the solar
energy to produce charge separation in a battery) have reached only
10-20\% efficiency. Thus our energy utilisation technology, particularly
at the molecular level, has much to learn from the design and functioning
of the photosynthetic apparatus.

\subsection{Some Biology}

Light harvesting complexes of photosynthetic organisms display a variety,
in terms of the pigment molecules involved and their structural arrangement.
The variations can be understood as the organisms' adaptations to the
specific living conditions, while the common features are likely to be
critical to the basic function. In particular, all light harvesting
complexes contain an antenna of pigment molecules that capture photons
and then direct the energy to the reaction centre where glucose is
synthesised. The advantages of such an arrangement are easy to understand:
\begin{itemize}
\item The pigment molecules are cheap, while the reaction centres are
expensive. The distribution of pigment molecules in an antenna therefore
increases the accessible light harvesting area at fixed cost.
\item The pigment molecules differ in their sensitivity to the light
spectrum and their orientations in the three dimensional antenna structure.
That increases the efficiency of photon capture, given the natural
variations in light direction, intensity and frequency.
\item Dissociation of water requires energy of several photons of visible
light that must be captured within a certain time. Light harvesting by an
antenna permits accumulation of the captured energy at the reaction centre.
\item The pigment molecules have a dead time after photon absorption.
They also switch off when the light is too bright. These properties are
easily accommodated by the antenna structure without affecting the goings
on at the reaction centre.
\item The pigment molecules are coupled to their neighbours via a rigid
protein scaffold, and the reaction centre is next to one of the pigment
molecules. That allows multiple transport pathways for the captured energy
to reach the reaction centre.
\end{itemize}

\begin{figure}
\epsfxsize=8cm
\epsffile{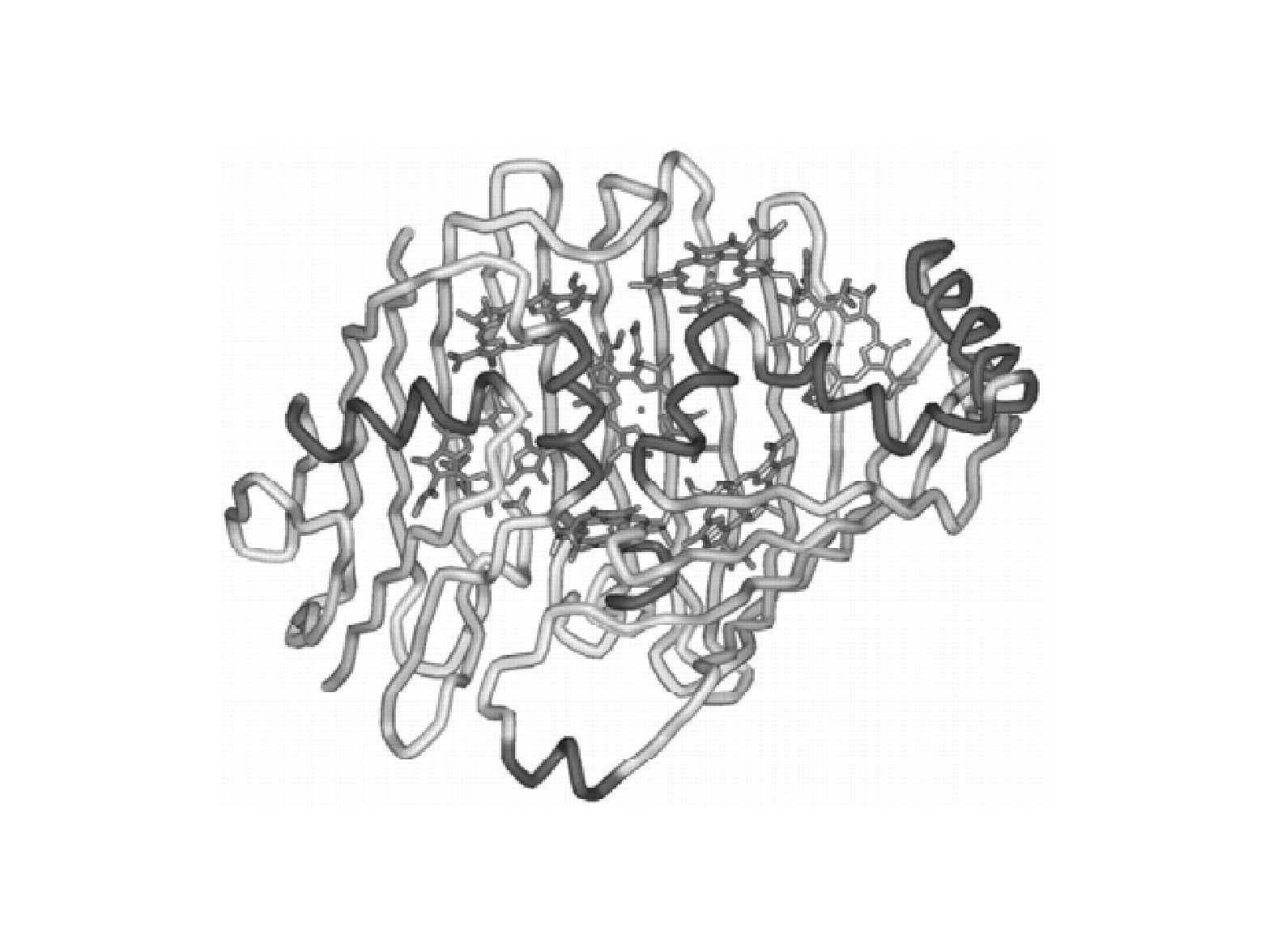}
\epsfxsize=8cm
\epsffile{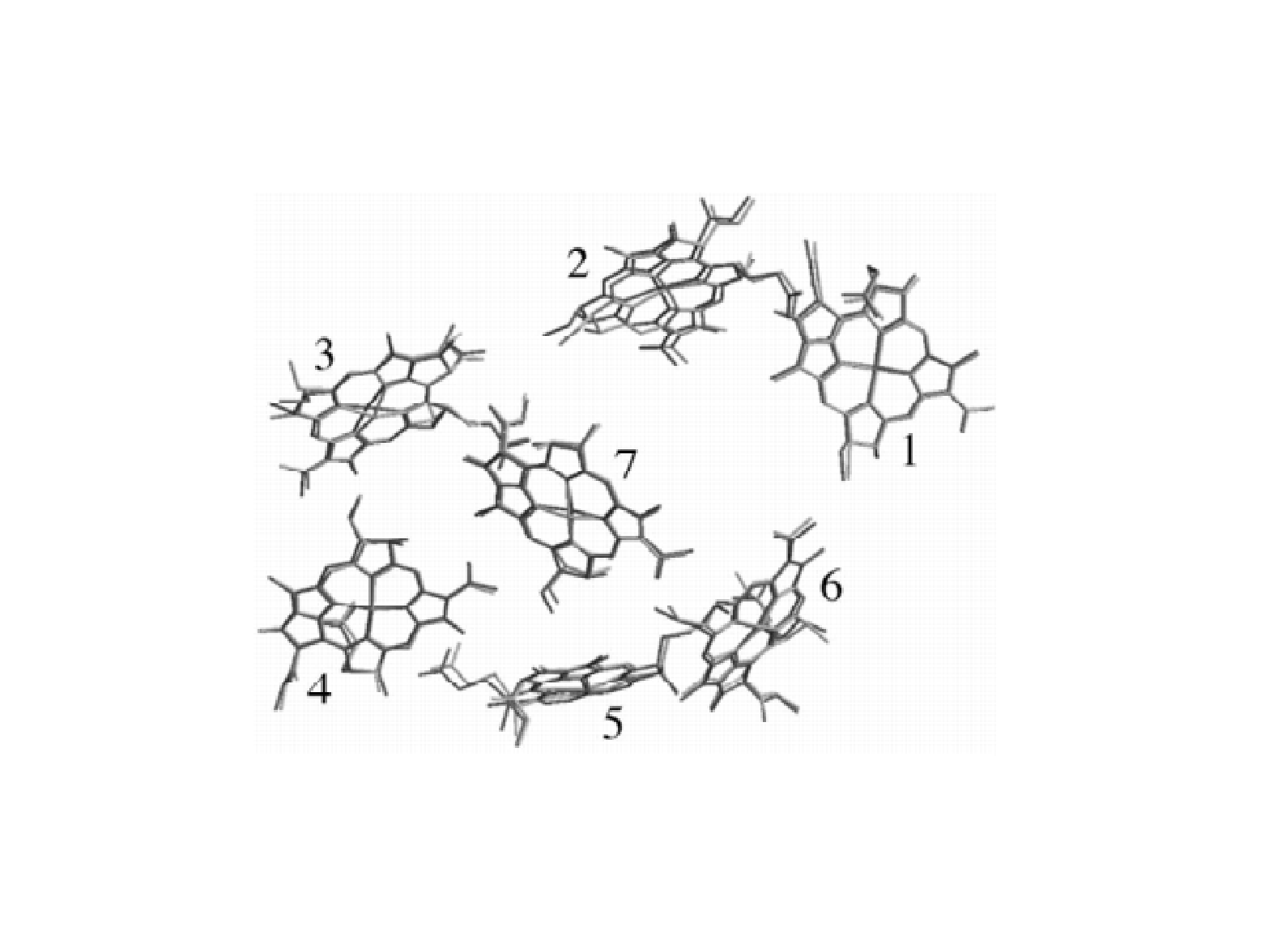}
\caption{The Fenna-Matthews-Olson antenna complex, with the protein scaffold
in the ribbon representation (left), and only the pigment molecules without
the scaffold (right). The reaction centre is next to pigment molecule 3.
The major energy transport pathways are $6-5-7-4-3$ and $1-2-7-3$. The figure
is reproduced from Ref.\cite{FMO2.2A}.}
\end{figure}

As an example, the Fenna-Matthews-Olson antenna complex of the purple
bacteria is shown in Fig.1. It contains 7 bacteriochlorophyll pigment
molecules, with neighbouring distances around 20\AA. The geometry is not
that of a star or a ring or some other regular graph, and the reaction
centre is not in the middle. The energy transfer time between neighbouring
pigment molecules is observed to be about 50fs.

Experimental investigations of the dynamics of energy transport in the
light harvesting complexes have thrown up a surprise. The process shows
wavelike coherent oscillations, going back and forth and lasting longer
than 400fs, which is quite distinct from classical random walk hopping
between neighbouring pigment molecules. The observations were made first
at liquid nitrogen temperature for purple bacteria containing the FMO
complex \cite{Fleming}, and then at ambient temperature for marine
cryptophyte algae containing PC645 and PE545 antenna proteins with
8 bilin molecules \cite{Scholes}.

The experiments are in the framework of two-dimensional spectroscopy.
They have used lasers on light harvesting complexes extracted from cells,
and not sunlight on living cells, but that is not believed to alter the
results. First one laser of frequency $\omega_1$ injects energy into the
complex creating an exciton, and then after a time lapse $\tau$, a second
laser of frequency $\omega_2$ extracts the energy from the complex by
stimulated emission. The results are displayed as intensity correlations
in the $\omega_1-\omega_2$ plane, evolving as a function of $\tau$.
(A video of the results of Ref.\cite{Fleming} is available as supplementary
information to the online version of the paper.) The diagonal peaks in
the plot are trivial, while the off-diagonal peaks describe couplings
between different frequency modes. The results show several oscillations
in the off-diagonal signal, going on without appreciable decay till the
maximum value of $\tau$ explored. (The maximum value of $\tau$ was set
by theoretical estimates, which turned out to be incorrect, and hence
we do not really know for how long the oscillations persist.). The
long-lasting coherent coupling between excitonic modes is the feature
that is closely tied to efficient energy transport and that a realistic
model of the process must explain.

\subsection{Some Chemistry}

Basic chemistry involved in photosynthesis is well-known \cite{psReview}.
Since the visible light wavelength is much larger than the separation
between pigment molecules, several pigment molecules can capture a given
incident photon. Still, every photon that falls on a light harvesting
complex is not captured. When the photon is captured, it creates a state
called exciton. Excitons are extended molecular (not atomic) states,
often spanning two pigment molecules over nanometre distances. Various
excitons in a light harvesting complex do not have the same energy.
Nevertheless, coherent energy transfer occurs between them, as clearly
demonstrated by the two-dimensional spectroscopy experiments.

The pigment molecules are too far away from each other for their molecular
orbitals to overlap. They interact with each other through the long range
dipole-dipole interaction. Photon absorption produces a polarisation cloud,
without any ionisation. Coulomb forces propagate the polarisation from one
pigment molecule to its neighbour. The tight covalent binding of the pigment
molecules to the protein scaffold couples this propagation to vibrational
modes, and the energy pulse hops around in the antenna complex.

A simple classical strategy for directing energy from the antenna of
pigment molecules towards the reaction centre is to have a funnel-shaped
energy landscape. That requires maintenance of permanent energy gradients
which is expensive, and using one battery to charge another is certainly
counterproductive. Furthermore, a non-dissipative pulse spends the least
time near the potential minimum, and so needs to be trapped at the right
moment to be useful. Needless to say, energy transport in photosynthesis
does not follow this strategy.

A realistic strategy has to incorporate the fact that the exciton energy
accumulates at the reaction centre, with a wavelike transport and without
dissipation. Furthermore, the energy needs to be trapped at the reaction
centre for a duration long enough ($\sim$ few ps) for conversion to
chemical form. Algorithms that can support such features have been
discovered in the study of quantum computation.

\subsection{Some Computer Science}

A physical process evolves some initial state to some final state
according to some interaction inbetween. That can be considered a
computation, provided a suitable map can be constructed between physical
properties of the system and abstract mathematical variables. In such
conditions, different physical interactions give rise to different types
of computers, while efficiency and stability of these computers depend
on the nature of the physical interactions implemented.

The traditional computational paradigm implements discrete Boolean logic
using digital electronic circuits. It is an instance of classical particle
dynamics. Quantum computation considers unitary evolution of quantum
states in a Hilbert space. That includes properties of both particle and
wave dynamics, and hence is more powerful than classical computation.
In a similar vein, one can look at wave computation \cite{wavecomput},
which is different from Boolean logic computation but contained in quantum
computation. Analog variables in wave computation can superpose, interfere,
disperse etc., but without any entanglement.

Historically, waves have been widely used in communications, but their
properties have hardly been exploited in computation. Wave algorithms
can cover the same Hilbert space as quantum computers, and the two have
the same time and oracle complexity. The two differ, however, in their
space requirements and stability properties. Explicitly, an $N-$dim
Hilbert space can be realised using $N$ wave modes but only $\log_2N$
qubits. On the other hand, wave dynamics is much more robust against
environmental disturbances than quantum dynamics is. A practical computer
always needs a trade-off between minimisation of resources and minimisation
of errors. Thus a wave computer can be advantageous in situations, where
the problem size is modest, spatial resources are cheap and a quantum
computer would be fragile.

A change in physical implementation of a mathematical algorithm can
also alter its interpretation. Both quantum and wave computers evolve
amplitudes of their states, subject to conservation of the norm.
The square of an amplitude, $|A|^2$, gives the probability of a state
in quantum systems, while it gives the energy of a mode in wave systems.
In particular, an amplitude amplification algorithm describes a search
process in the quantum setting and an energy focusing process in the wave
setting. Search algorithms and their optimisation have been extensively
investigated in the framework of quantum computation. Their wave versions
are efficient schemes to transfer/redistribute energy with many practical
applications---from mechanical systems to chemical, electrical and
biological ones.

A wave dynamics realisation of Grover's algorithm is what a mechanic
intuitively does to find a structural defect in an object---give it
a vigorous shake. The vibrational modes go out of phase at the location
of the defect, which leads to amplitude and energy enhancement at that
place and subsequent cleavage. Parameters of the process, vibrational
frequencies and phase shift at the defect, can be optimised for the
best results. But even when the parameters are not perfectly tuned,
the algorithm works reasonably well providing useful results.

With this background, we can see that the energy transport problem in
photosynthesis maps to the spatial search problem, where an initial
amplitude distribution over many locations evolves by transfer between
neighbours and gets concentrated at a specific target location.
To ascertain whether that is indeed what nature has implemented,
we need to figure out:
(1) To what extent is the physical hardware of the light harvesting
complex capable of supporting the spatial search algorithm?
(2) Are there any experimentally testable characteristic signatures
indicating the role of wave dynamics in the optimal solution to the
problem?

\subsection{Some Physics}

Since the discovery of coherent energy transport in photosynthesis,
many attempts to model it on the basis of quantum dynamics have been
made (see for instance, Refs.\cite{psEntangl,psModel1,psModel2} and
references therein). These have analysed effects of entanglement,
decoherence and environmental noise in fully quantum systems. My goal
here is more modest---the process can be understood using classical
dynamics of waves, without invoking full-fledged quantum dynamics.

Any state in an $N-$dimensional Hilbert space can be decomposed as
$|\psi\rangle = \sum_{i=1}^N c_i |i\rangle$. Its evolution is called
coherent, when the relative phases of $c_i$ (corresponding to the
off-diagonal elements of the density matrix) are protected from external
disturbances. It is called entangled, when specific to a choice of
subsystems, the state cannot be expressed as a product of the subsystem
states. Thus coherence is a generic property arising from superposition,
while entanglement always refers to division of the system in to smaller
subsystems. A given Hilbert space can be realised in many ways depending
on the physical components involved. When it is a direct sum of distinct
modes (i.e. $N=1+1+\ldots+1$), it exhibits superposition without any
entanglement. When it is a tensor product of smaller spaces (i.e.
$N=N_1\times N_2$), entanglement is possible. It is the former realisation
that appears in wave computation, while the latter one is routinely used
in quantum computation.

In the real world, all wave processes suffer decoherence and damping due
to interaction with the environment. But it is observed that decoherence
is much more rapid than damping (i.e. reduction in the norm of the state),
and superposition is much more stable than entanglement. These properties
make classical wave algorithms much more stable against environmental
disturbances than their quantum counterparts. Indeed, a variety of wave
systems with coupled vibrational/rotational modes and small damping can
be constructed easily.

With the above concepts in mind, we can list the physical ingredients
that can produce long-lived coupled wave modes in the light harvesting
antenna, as the hardware for implementing the energy transport algorithm:
\begin{itemize}
\item The dipole-dipole interaction leads to vibrational motion of the
polarisation cloud. The charge density degree of freedom (i.e. $\rho=
\psi^\ast\psi$) does not involve the electronic wavefunction phase that
decoheres rapidly. But it does carry the more robust classical vibrational
phase.
\item Classical wave modes belong to the ``decoherence free" subspace
of the full quantum space. They correspond to coherent states, which
are eigenstates of the annihilation operator and not energy eigenstates.
Then, by definition, they remain unaffected by linear coupling to the
environment.
\item Decoherence rates are often estimated from system-environment
weak scattering cross-sections, dilute gas approximation and Fermi's
golden rule. Such calculations generically overestimate decoherence
\cite{Leggett,Unruh}. In particular, the estimates are invalid for
frequent interactions (what is supposed to escape irreversibly actually
returns), as well as for adiabatic interactions (slow phase disturbances
cancel in case of cyclic evolutions). As a result, only environmental
modes with evolution time scales comparable to that for the system
contribute significantly to decoherence of waves. In this context,
it is worth noting that the photon energy being transported during
photosynthesis is considerably larger than the energy of the thermal
phonon disturbances.
\end{itemize}

\section{A Simple Model}

Having described the essential requirements of an efficient energy
transport algorithm, we now look at an explicit model of it involving
wave computation. The spatial search algorithm \cite{hypsrch} is a
variation on Grover's database search algorithm, where locality of
physical interactions limits the movement during the search process
from a location to only its neighbours. Geometry and connectivity
of the distributed database are important features in the process.
The algorithm needs two basic operations: amplitude transfer from any
location to its neighbours and an amplitude reflection oracle at the
target location. The optimal solutions require database exploration
with relativistic dispersion relation, which is satisfied by wave
propagation but not by diffusion.

A model of coupled simple harmonic oscillators that can implement
the algorithm is displayed in Fig.2. The $N$ identical oscillators
representing the pigment molecules are shown as a linear chain, but
they can also be arranged in a network with suitable connectivity.
The reaction centre is represented as a different oscillator on a
side branch. The nearest neighbour coupling among the oscillators
allows amplitude transfer with relativistic dispersion relation.
Reflection of a wave pulse from a hard wall flips the amplitude
in sign, as depicted in Fig.3 using the method of images. For a
propagating wave pulse, the connection to the side branch acts as
a beam-splitter, where reflection as well as transmission occur.
The frequencies of the oscillators are the parameters that have to
be optimised, together with the geometry of the oscillator network.

It is important to note that in the optimisation of biological resources,
as exemplified by Darwinian evolution, energy is more important than time
(living organisms slow down their metabolism in adverse environments),
and time is more important than space (parallel processing is commonplace).
These criteria differ from those normally used in studying computational
complexity. In the present case, the dynamics of the oscillators is to be
optimised for the most efficient energy transfer, and not for the quickest
energy transfer.

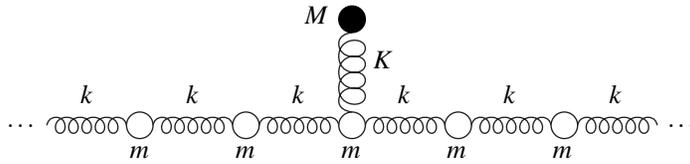
\begin{figure}
\begin{picture}(230,60)(0,0)
\Text(-10,10)[c]{$\ldots$}
\Gluon(0,10)(30,10){3}{5}    \Text(15,22)[c]{$k$}
\BCirc(35,10){5}             \Text(35,0)[c]{$m$}
\Gluon(40,10)(70,10){3}{5}   \Text(55,22)[c]{$k$}
\BCirc(75,10){5}             \Text(75,0)[c]{$m$}
\Gluon(80,10)(110,10){3}{5}  \Text(95,22)[c]{$k$}
\BCirc(115,10){5}            \Text(115,0)[c]{$m$}
\Gluon(120,10)(150,10){3}{5} \Text(135,22)[c]{$k$}
\BCirc(155,10){5}            \Text(155,0)[c]{$m$}
\Gluon(160,10)(190,10){3}{5} \Text(175,22)[c]{$k$}
\BCirc(195,10){5}            \Text(195,0)[c]{$m$}
\Gluon(200,10)(230,10){3}{5} \Text(215,22)[c]{$k$}
\Text(240,10)[c]{$\ldots$}

\Gluon(115,15)(115,45){5}{4} \Text(127,35)[c]{$K$}
\GCirc(115,50){5}{0}         \Text(102,52)[c]{$M$}
\end{picture}
\caption{A linear chain of identical harmonic oscillators, coupled to
a side branch oscillator that acts as a storage cavity.}
\end{figure}

\begin{figure}
\epsfxsize=8cm
\epsffile{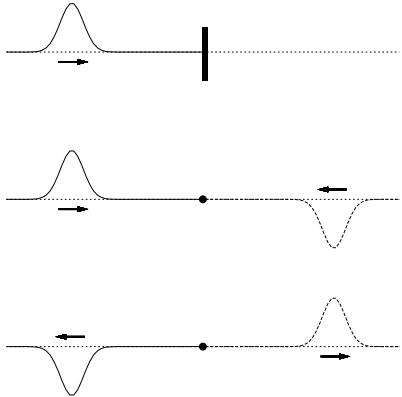}
\caption{The reflection oracle for a wave pulse incident on a hard wall.
The node at the wall (top) is simulated using the method of images
(middle), and the linear superposition evolves to have the amplitude
of the reflected pulse reversed in sign (bottom).}
\end{figure}

The mathematical algorithm that does this job was written down by Tulsi
\cite{Tulsi}. It works in an $(N+1)-$dimensional Hilbert space, with
the extra state playing the role of a trapping cavity at the target
location. The amplitude evolves essentially in the three-dimensional
subspace of the complete Hilbert space, formed by the cavity state,
the target state coupled to the cavity and an ``averaged" non-target
state. Clever transfer of amplitude between the distributed database
and the cavity allows $O(1)$ of the initial energy to accumulate in the
cavity. That amounts to a resonance condition: For constructive build up
of the amplitude, the phase change for the energy pulse to cycle in the
oscillator network has to match the phase change in the side branch.
(Note that a similar strategy is used in gravitational wave detectors,
where a weak vibrational signal accumulates in a high$-Q$ Fabry-P\'erot
cavity, until it crosses the detection threshold.) Indeed, the resonance
condition is a prediction of the algorithm that can be experimentally
tested as a requisite property of the light harvesting complex.

The experimentally studied light harvesting complexes have a small
number of pigment molecules, seven for the purple bacteria and eight
for the marine cryptophyte algae. In such situations, instead of
looking at the asymptotic behaviour of the algorithm, one should look
for specific values of $N$ and the corresponding connectivity of the
oscillators that yield high energy accumulation efficiency. (Such
efficient solutions are well-known for Grover's database search
algorithm.) Detailed analysis and results will be presented elsewhere.


\begin{theacknowledgments}
I am grateful to Graham Fleming, Anthony Leggett and K.L. Sebastian
for useful discussions. I thank the audience for their many comments
and questions that helped me clarify the concepts presented here.
\end{theacknowledgments}


\begin{thebibliography}{9}

\bibitem{psReview}
R. van Grondelle \emph{et al.},
\emph{Biochimica et Biophysica Acta} \textbf{1187}, 1--65 (1994).

\bibitem{FMO2.2A}
A. Camara-Artigas R.E. Blankenship and J.P. Allen,
\emph{Photosynth. Res.} \textbf{75}, 49--55 (2003).

\bibitem{Fleming}
G.~S. Engel \emph{et al.},
\emph{Nature} \textbf{446}, 782--786 (2007).

\bibitem{Scholes}
E. Collini \emph{et al.},
\emph{Nature} \textbf{463}, 644--647 (2010).

\bibitem{wavecomput}
A. Patel,
\emph{Int. J. Quant. Inform.} \textbf{4}, 815--825 (2006);
Erratum \emph{ibid.} \textbf{5}, 437 (2007).

\bibitem{psEntangl}
K.~B. Whaley, M. Sarovar and A. Ishizaki,
in \emph{``Quantum Effects in Chemistry and Biology"},
Proceedings of the 22nd Solvay Conference on Chemistry,
e-print arXiv:1012.4059 (2010).

\bibitem{psModel1}
F. Caruso \emph{et al.},
\emph{Phys. Rev. A} \textbf{81}, 062346 (2010).

\bibitem{psModel2}
V. Vedral and T. Farrow,
e-print arXiv:1006.3775 (2010).

\bibitem{Leggett}
A.~J. Leggett, in \emph{Quantum Mechanics of Complex Systems, I and II},
Proceedings of the 1989 NATO Advanced Study Institute, Evora, Portugal,
Plenum Press, New York, 1990.

\bibitem{Unruh}
W.~G. Unruh and W.~H. Zurek,
\emph{Phys. Rev. D} \textbf{40}, 1071 (1989).

\bibitem{hypsrch}
N. Shenvi, J. Kempe and K.~B. Whaley,
\emph{Phys. Rev. A} \textbf{67}, 052307 (2003).

\bibitem{Tulsi}
A. Tulsi,
\emph{Phys. Rev. A} \textbf{78}, 012310 (2008).

\end{thebibliography}
\end{document}